\documentclass[english,aps,reprint,showpacs,titlepage,longbibliography]{revtex4-2}

\usepackage[T1]{fontenc}	
\usepackage{array}
\usepackage{tabularx}
\usepackage{geometry}
\usepackage{graphicx}
\usepackage{times}
\usepackage{hyperref}  
\hypersetup{colorlinks=true,urlcolor=blue,citecolor=blue,linkcolor=blue}
\usepackage{array}
\usepackage{enumerate}
\usepackage{enumitem}  
\usepackage{amsmath}
\usepackage{amssymb}
\usepackage{tikz}
\usepackage{graphicx}
\usepackage{multirow}
\usepackage{tcolorbox}
\usepackage{ragged2e}
\usepackage{tcolorbox}
\usepackage{float}
\usepackage[utf8]{inputenc}

\usepackage[normalem]{ulem}
\usepackage{placeins}
\usepackage{parskip}

\begin{document}
\begin{titlepage}

\title{Game-Based vs. Simulation-Based Instruction: exploring the sequencing effect on elementary pre-service teachers' understanding of the photoelectric effect}

 \author{Razan Hamed}
 \affiliation{Department of Physics and Astronomy, Purdue University, 525 Northwestern Ave, West Lafayette, IN, 47907, U.S.A.}

 \author{N. Sanjay Rebello}
 \affiliation{Department of Physics and Astronomy, Purdue University, West Lafayette, IN, 47907, U.S.A.\\
 Department of Curriculum and Instruction, Purdue University, West Lafayette, IN, 47907 U.S.A.} 
 

\begin{abstract}

The use of digital tools and multiple representations like educational games and interactive simulations is of great importance to physics education. This study investigates the sequencing effects of an educational video game ‘Photon Jump’ and the PhET Photoelectric Effect simulation on pre-service teachers’ understanding of the photoelectric effect. Using a counterbalanced quasi-experimental crossover design, pre-service teachers enrolled in a physics course (N = 83) experienced both interventions in opposite orders. Conceptual understanding was measured across three standardized assessments, complemented by open-ended reflection questions on participants’ preferences and willingness to use both tools for future learning. The simulation-first sequence yielded a greater significant improvement in performance $p = 0.001$ as compared with game-first sequence $p = 0.06$. participants' preferences for using the game as opposed to the simulation were dependent on the sequence that they were randomly assigned to. Findings underscore the complementary strengths of game-based and simulation-based instruction, highlighting the importance of choosing the right sequence when using multiple representations in teaching abstract physics phenomena to pre-service teachers.

\clearpage
\end{abstract}
\maketitle
\end{titlepage}

\section{Introduction}
\label{sec:intro}

In recent years, educational video games and interactive simulations have gained increasing traction as digital tools to enhance conceptual learning in science education. The photoelectric effect - a phenomenon foundational to quantum physics - is one such concept that participants frequently find challenging \cite{steinberg1996development}. Research shows that coordinating multiple, well-designed representations improves comprehension and transfer of learning especially for complex science topics \cite{ainsworth2006deft}. Previous studies have explored using multiple representations but have not investigated the impact of the order in which such representations are introduced to the participants. This study explores the sequencing effect and the implications of using an educational video game ‘Photon Jump’ along with a PhET simulation, on pre-service teachers’ understanding of the photoelectric effect.
 We aim to answer the following research questions:  How does the sequence of instructional interventions (simulation followed by game vs. game followed by simulation) influence participants'...
\begin{itemize}
    \item [\textbf{RQ 1.}] conceptual understanding of the photoelectric effect, as measured by assessment scores? 
    \item [\textbf{RQ 2.}] preferences for each instructional tool and  willingness to use each tool in future learning?
\end{itemize}

\section{Background}
\label{sec:background}

Physics education research has always emphasized the importance of active, student-centered instructional approaches for improving conceptual understanding and engagement among learners \cite{hake1998interactive}. Educators and researchers in this field have explored many alternative learning tools like educational games and simulations and their impact on participants learning and perceptions of physics  \cite{richter2025gamification, banda2021effect, pranata2024physics}. 

Physics simulations typically provide interactive visualizations of physical phenomena based on accurate mathematical models that allow learners to manipulate variables and observe real-time effects helping them explore otherwise abstract or sometimes inaccessible scenarios \cite{de1998scientific, rutten2012learning}. Well-known platforms like PhET Simulations have demonstrated consistent success in supporting inquiry-based learning and facilitating exploration of physical relationships in fields like mechanics, circuits, and even introductory quantum mechanics \cite{perkins2006phet}. Research shows that simulations can be particularly effective in correcting misconceptions when coupled with guided inquiry \cite{zacharia2011physical, adams2008study}. In this study, simulations will be referred to as 'direct representations' of physical phenomena since they include elements that map directly on such how such phenomena happen in real life. 

Educational physics games typically incorporate challenge, feedback, and exploration into physics learning using metaphors, allegories, and artificial constraints that do not always mirror exact physical systems but rather invite conceptual transfer and scaffolding \cite{gee2005video, clark2016digital}. Educational games are praised for their motivational appeal and capacity for sustained engagement especially for participants who might lack internal motivation \cite{squire2013video, papastergiou2009digital}. They promote situated learning in which learners acquire knowledge through contextualized activity and feedback loops embedded within a game system \cite{barab2009educators}. In this study, educational games will be referred to as 'analogical representations' inspired by Gentner’s Structure-Mapping Theory, which states that analogical reasoning enables learners to map relationships from a familiar or fictional domain onto a target concept \cite{gentner1983structure}. 

When considering the type of instructional tool to use for teaching physics, it is important to consider student's prior physics preparation as well as their self-efficacy and attitudes toward physics, as these factors can significantly affect their learning outcomes \cite{hazari2007gender}. Research shows that learning physics can be especially intimidating for pre-service teachers due to physics anxiety and low self-efficacy that shows more often in female than male future teachers \cite{ccalicskan2017physics}. Many pre-service teachers enter their teacher education programs with limited background knowledge in physics and math, and in many instances, with a negative attitude towards physics, often caused by previous learning experiences they might have had in the past \cite{menon2016preservice}. Research also shows that low confidence and physics anxiety can reduce pre-service teachers' motivation and engagement in physics courses leading to lower learning outcomes \cite{ccalicskan2017physics}.

However, despite these challenges, physics remains an important component of teacher science education. A strong understanding of physics helps pre-service teachers develop critical thinking, problem-solving, and scientific reasoning skills that are essential for effective teaching \cite{sadaghiani2010scientific}. Therefore, it is crucial to support pre-service teachers in learning physics using innovative instructional tools that lower their anxiety and increase their motivation. Examples of such instructional tools are games and simulations which have been studied thoroughly in literature. 

Several studies highlight the complementary nature of games and simulations. For example, Rieber \cite{rieber1996seriously} showed that games and simulations can enhance conceptual understanding, especially for novice learners, as they actively build their knowledge rather than passively receive information. Along with Rieber \cite{rieber1996seriously}, Bressler \cite{bressler2013mixed} found that combining metaphorical game play with simulation environments improved both engagement and conceptual gains compared to either one alone. Nevertheless, no known studies have explored the effect of sequencing games and simulations in physics learning. This study contributes to the existing literature by exploring how the sequencing of games and simulations affects pre-service teachers learning outcomes with respect to understanding abstract physics concepts, particularly the photoelectric effect.

\begin{figure*}
    \centering
    \includegraphics[width=1\linewidth,height=5cm]{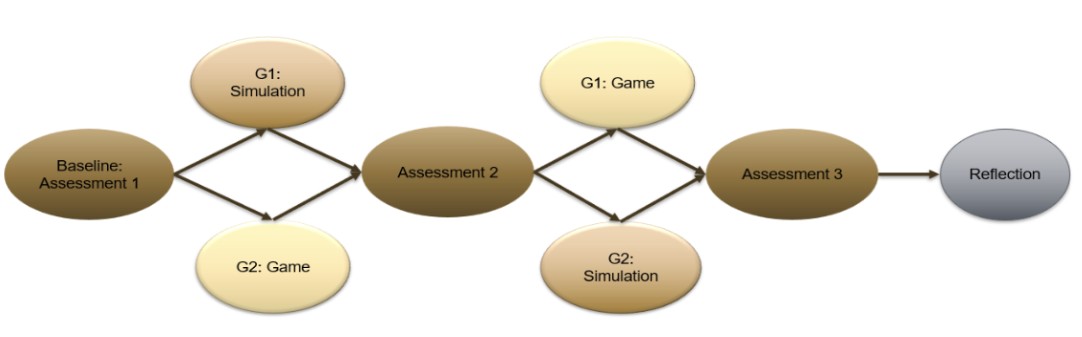}
    \caption{Study Design}
    \label{Fig:StudyDesign}
\end{figure*}

\begin{figure*}
        \centering
        \includegraphics[width=1\linewidth,height=5cm]{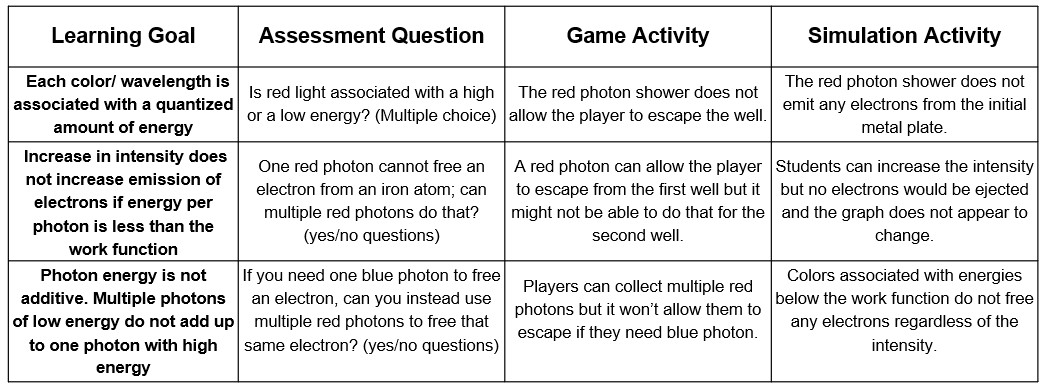}
        \caption{Assessment Design}
        \label{Fig:AssessmentDesign}
    \end{figure*}

\section{Methods}
\label{sec:methods}
\subsection{Research Design}
This study was conducted in a physics course for elementary education majors at a large public Midwestern university. The course consisted of 93\% female and 7\% male pre-service teachers taking this course to complete their education degree. A total of $N = 83$ participants were divided into two groups of comparable sizes with similar levels of physics competency. Each of the two groups was assigned to a distinct instructional condition. Both conditions introduced the same instructional modalities - a game-based activity and a simulation-based activity - but in reverse order to investigate the effects of intervention sequencing on pre-service teachers' learning outcomes. 

This study employed a counterbalanced quasi-experimental crossover design consisting of two instructional interventions (game and Simulation) and three assessments (pre-test, mid-test, and post-test) as shown in Fig \ref{Fig:StudyDesign}. The three assessments were standardized in format, content, and difficulty and were mapped onto specific learning objectives and unique activities within both the game and the simulation as shown in Fig \ref{Fig:AssessmentDesign}.

All participants first completed an identical baseline assessment (pre-test) to check for prior knowledge regarding the photoelectric effect, then engaged in their first assigned intervention - simulation for Group 1 and game for Group 2. Following that, all participants completed a second assessment (mid-test) which was implemented to check for any increase in scores based on the first intervention alone. Following the second assessment, participants completed the second intervention - game for Group 1 and simulation for Group 2. Lastly, all participants completed the third assessment (post-test) to measure the effectiveness of the second intervention and to establish a final score for each sequence of interventions. participants in each group worked individually to complete the activities and were provided with a worksheet to guide them through each activity. Participants also responded to five open-ended reflection questions regarding the game and the simulation to share their experiences and preferences regarding both learning modalities. Participants completed the questions individually and their responses were analyzed quantitatively using excel and python. Correct responses to the three assessments were tallied to produce an overall performance score which was used to assess the instructional effectiveness of the two interventions. Participants’ reflections were collected and analyzed qualitatively. 

\subsection{Instructional Interventions}    
\subsubsection{‘Photon Jump’ Photoelectric Effect Game} 
\begin{figure}[h!]
        \centering
        \includegraphics[width=0.48\textwidth,height=4.8cm]{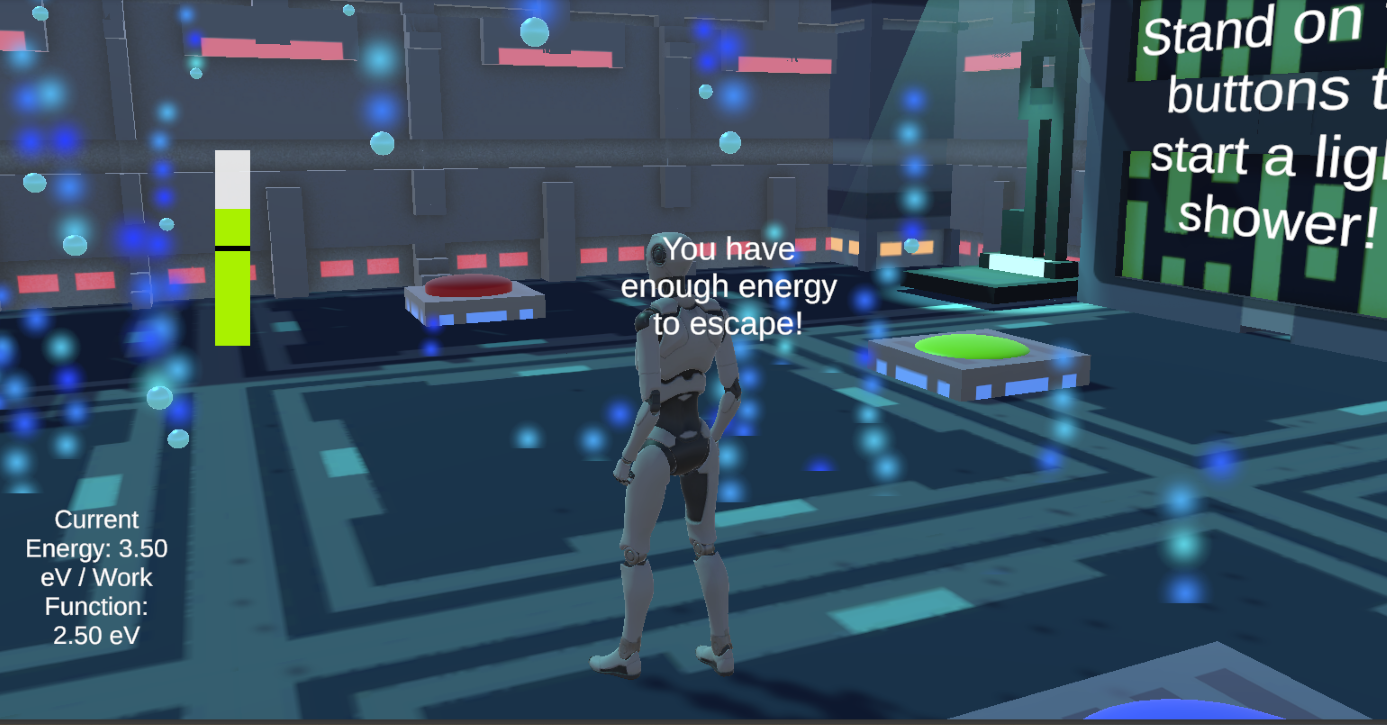}
        \caption{A screenshot of the 'Photon Jump' Game}
        \label{Fig:PhotonJumpGame}
    \end{figure}

 The primary tool for this activity was an interactive 3D video game titled “Photon Jump” that serves as an analogical representation of the photoelectric effect to enhance pre-service teachers' conceptual understanding of such phenomenon. In “Photon Jump” (Fig. \ref{Fig:PhotonJumpGame}), the player assumes the role of an electron trapped inside a four-walled room with a goal to jump over a wall to escape the room by gaining the right amount of energy from incoming photons. Before receiving that energy, the player would not be able to jump. The player's escape resembles the ejection of an electron from the surface of a metal once that electron absorbs a photon's energy that is at or above the work function. To receive incoming photons, players have to activate a photon 'shower' resembling a photon beam by standing a one of the colored buttons. Players have to try different buttons to find the right photon color that corresponds with the right energy. Players receive instructions and real-time feedback in the game on whether they have absorbed the right amount of energy to be able to escape. 
 
 Once players escape the room, they come across a timed obstacle course where they have to jump over barriers and avoid falling into the abyss in order to reach a light bulb as fast as possible and turn it on. Lighting up the bulb resembles a current of electrons that acquired some kinetic energy after being ejected from the metal. Participants played the game individually and answered worksheet questions as they tried different maneuvers in the game. The worksheet prompted the participants to connect the game elements with specific physical elements in the photoelectric effect phenomenon. 

\subsubsection{PhET Photoelectric Effect Simulation} 
\begin{figure}[h!]
        \centering
        \includegraphics[width=0.48\textwidth,height=4.8cm]{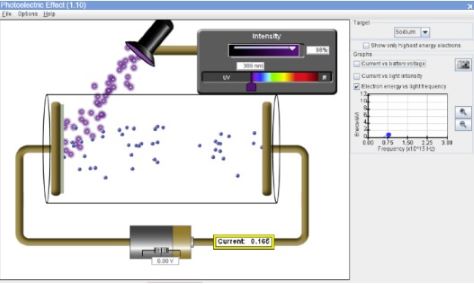}
        \caption{The 'Photoelectric Effect' PhET simulation}
        \label{PhETSimulation}
    \end{figure}

(Fig. \ref{PhETSimulation})
The “Photoelectric Effect” simulation - accessible via https://phet.colorado.edu/en/simulations/photoelectric - served as a direct representation of the photoelectric effect where participants were able to manipulate different variables and parameters such as light wavelength and intensity and observe their effects on the photoelectron emission or lack thereof from a metal plate's surface. Along with the simulation itself, participants were provided with a structured lab worksheet that guided them through different aspects of the simulation and prompted them to do simple analytical calculations regarding energy and frequency.

Analogous to the game, the simulation included a visual display of a photon shower/beam and a stream of electrons ejected from the left plate once they absorb the right amount of energy. Moreover, participants observed and analyzed a graph of frequency vs energy illustrating the linear relationship between the two quantities once the energy threshold or the work function is met.

\section{Results}
\label{sec:results}

\begin{figure}[h!]
        \centering
        \includegraphics[width=0.48\textwidth,height=4.8cm]{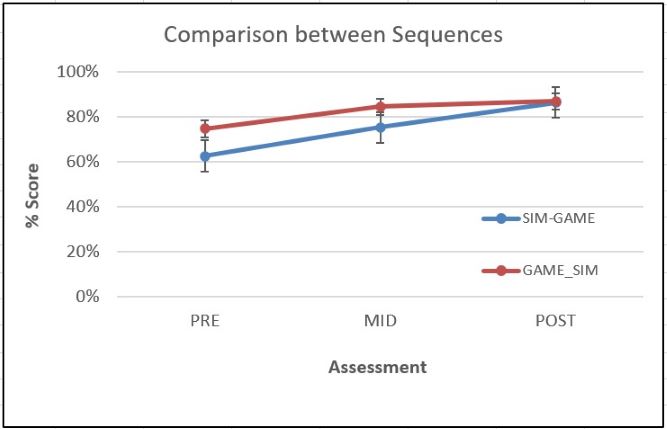}
        \caption{Effect of sequencing on assessment scores. The error bars represent standard error.}
        \label{Fig:Results}
    \end{figure}

In this study, participants completed three assessments; a base-line assessment (pre-test) before the first intervention, a second assessment (mid-test) after the first intervention, and a third assessment (post-test) after the second intervention. The results of the first assessment, as shown in Fig. \ref{Fig:Results}, indicate that the participants had some prior knowledge about the photoelectric effect which explains the high scores in the pre-test. The scores of all three assessments were calculated and analyzed through excel and a python code. The pre-test scores averaged at 62.7\%  for the first group and 74.8\% for the second. The mid-test scores for the first and second groups were 75.4\% and 84.6\% respectively. Lastly, the post-test scores were 86.5\% and 87.0\% respectively. While the post-test scores for both groups are comparable, the growth rate in each group was different.   

A single factor ANOVA showed that the difference between the pre-test and the post-test was statistically significant $p = 0.001$ in the first group (Simulation-Game) but not significant $p = 0.06$ in the second group (Game-Simulation). This results indicates a meaningful sequencing effect in the two conditions. Notably, the starting score for the Simulation-Game group was lower than that of the Game-Simulation group and could be mistaken as the reason behind the statistically significant difference in the first group. Therefore, two moderation regression analyses were conducted to examine whether the effect of pre-test scores on post-test scores was different depending on the condition participants were in (game-simulation vs simulation-game). 

In the first analysis, pre-test scores were used to predict mid-test scores. The results showed that participants who scored higher on the pre-test also tended to score higher on the mid-test. However, the interaction between pre-test scores and condition was not statistically significant, $p=.241$, indicating that the relationship between pre-test and mid-test performance did not differ significantly across conditions. 

In the second analysis, mid-test scores were used to predict post-test scores. Again, participants who scored higher on the mid-test tended to score higher on the post-test. However, in this case, the relationship between mid-test and post-test scores was different across the two conditions. The interaction between mid-test scores and condition was statistically significant, $p=.039$ indicating that the association between mid-test and post-test performance differed across the two conditions. This suggests that the condition influenced how participants’ performance changed from the mid-test to post-test.  

In addition to assessment scores, participants’ reflections were also collected and analyzed revealing aspects that participants found most beneficial and most challenging about both interventions. In the simulation, many participants liked the ability to manipulate photon wavelengths and intensity levels whereas in the game they enjoyed the challenge of finding the right color of photons to escape and the ability to make mistakes and try again. The majority of participants indicated that they found the game easy to play despite many being new to playing video games which shows that lack of prior gaming experience did not affect the participants performance in the game.

The participants' preferences for the game and the simulation where mixed across the two groups and was influenced by the sequence that each group was assigned to. Group 1 (simulation-game) were split almost evenly with some preferring the game and other preferring the simulation whereas participants in Group 2 (game-simulation) tended to favor the game over the simulation. Nevertheless, a majority of participants in both groups said they would play the game again or a similar game in the future and indicated that they would be open to using such game in their own classrooms to teach their future elementary students about simple science concepts. 
\setlength{\parskip}{0pt}
\section{Discussion and Conclusion}
\label{sec:discussion}

This study provides evidence for the complementary strengths of game-based and simulation-based learning in enhancing conceptual understanding of the photoelectric effect. It specifically highlights that While both modalities effectively support student learning, the order in which each modality is presented has significant impact on pre-service teachers' learning outcomes as measured by the three assessments. The simulation-game sequence has proven to be more effective for student's learning than the game-simulation sequence. Although each group had a different baseline score in the initial assessment, moderated regression analyses showed that the pre-test scores were not the reason for the increase in scores later on. This result confirms that the condition/order of interventions was the underlying factor behind the change in participants' post-test scores. The order of intervention also seemed to influence participants' preferences for using either the game or the simulation; the majority of participants whose first intervention was the game preferred the game, whereas for those whose first intervention was the simulation did not have a clear favorite in their preference for either the game or the simulation. 

\vspace{5 mm}

These findings emphasize the importance of thoughtful integration of diverse forms of representation (direct and analogical) in physics teaching in order to maximize learning and engagement especially with complex topics like the photoelectric effect. By leveraging the unique strengths of each modality, educators can capture students' attention and spark their curiosity through the game while providing precise parameter-controlled observations through the simulation. Future work will build on the findings of this study by integrating AI chat bots that would serve as tutors for pre-service teachers while navigating through the game and the simulation.

\section{Acknowledgments}
ChatGPT-4.0 was employed for improving word choices and sentence structure and for generating code that was used for the statistical analysis of the data. 
\clearpage
\bibliography{references}

\end{document}